\documentclass[11pt]{cernrep}
\usepackage{graphicx}
\usepackage{here}
\begin{document}
\def\tb{\tilde{b}}

\title{$Z^0$ and $W$ transverse momentum spectra at the LHC}

\author{Xiaofei Zhang and George Fai}
\institute{Center for Nuclear Research, Department of Physics,
Kent State University \\
Kent, Ohio 44242, USA}
\maketitle


At LHC energies, perturbative QCD (pQCD) provides a powerful 
calculational tool. 
For $Z\ ^{0}$  and $W$   transverse momentum spectra, 
pQCD theory  agrees with the CDF\cite{CDF-Z} and D0 \cite{D0-W} data 
very well
at  Tevatron energies\cite{qz01}.  The LHC $pp$  program will test 
pQCD at an unprecedented  energy.   
The heavy-ion program at the LHC will make it  possible for the first time to 
observe the full $p\ _T$ spectra of heavy vector bosons in nuclear
collisions and will provide a testing ground for pQCD resummation theory
\cite{CSS-W}.

In nuclear collisions, the power corrections 
will be enhanced by   initial and final state multiple scattering. 
As we  will show, the high-twist effects are  small at LHC 
for heavy boson production.  The only important nuclear
effect  is the nuclear modification of the parton 
distribution function (shadowing).
Because of  their large masses,  $W$ and $Z^0$ will 
tell us about  the nuclear Parton Distribution Function (nPDF) at large 
scales.  
Since  for the $p\ _T$  spectra of heavy bosons 
contributions from different scales need to be resummed, 
the  $p\ _T$ spectra for heavy vector bosons 
will  provide  information 
about the evolution of  nPDFs from small scales to large scales.
Since it is more difficult to detect $W^\pm$ than 
$Z^0$, we will concentrate on discussing  $Z^0$ here
(the results
for $W^{\pm}$ production are very similar\cite{zhang-fai}).

Resummation of the large logarithms in QCD can be carried out either
in $p\ _T$-space directly, or in the so-called 
``impact parameter'', 
$\tb$-space, which is a Fourier conjugate of the $p\ _T$-space.
Using the renormalization group
equation technique, 
Collins, Soper, and Sterman (CSS) derived a formalism
for the transverse momentum distribution of vector boson production
in hadronic collisions\cite{CSS-W}. In the CSS formalism, non-perturbative
input is needed for the large \ $\tb$ region.  The dependence of the pQCD
results on the non-perturbative input is not weak if the original
extrapolation proposed by CSS is used. Recently, a new extrapolation
scheme was introduced, based on solving the renormalization group equation
including power corrections\cite{qz01}. Using
the new extrapolation formula, the dependence of the pQCD results
on the non-perturbative input was significantly reduced.

For vector boson ($V$) production in a hadron collision, 
the CSS resummation formalism yields\cite{CSS-W}:
\begin{eqnarray}
\frac{d\sigma(h_A+h_B\rightarrow V+X)}{dM^2\, dy\, dp_T^2} =
\frac{1}{(2\pi)^2}\int d^2 \tb\, e^{i\vec{p}_T\cdot \vec{\tb}}\,
\tilde{W}(\tb,M,x_A,x_B) + Y(p_T,M,x_A,x_B) \,\,\, ,
\label{css-gen}
\end{eqnarray}
where $x_A= e^y\, M/\sqrt{s}$ and $x_B= e^{-y}\, M/\sqrt{s}$, with
rapidity $y$ and collision energy $\sqrt{s}$.
In Eq.~(\ref{css-gen}), the 
$\tilde{W}$ term dominates the $p\ _T$ distributions
when $p\ _T \ll M$, and the $Y$ term gives corrections 
that are negligible
for small $p\ _T$, but become important when $p\ _T\sim M$. 

The function $\tilde{W}(\tb,M,x_A,x_B)$ can be  
calculated perturbatively for small \ $\tb$, but an
extrapolation to the large $\tb$ region requiring nonperturbative input
is necessary in order to complete the Fourier transform in Eq.~(\ref{css-gen}).
In oder to improve the situation, a new form was proposed\cite{qz01}
by solving the renormalization equation including power
corrections. In the new formalism, 
$\tilde{W}(\tb,M,x_A,x_B)=\tilde{W}^{pert}(\tb,M,x_A,x_B)$, when 
$\tb \leq \tb_{max}$, with 
\begin{equation}
\tilde{W}^{pert}(\tb,M,x_A,x_B) =
{\rm e}^{S(\tb,M)}\, \tilde{w}(\tb,c/\tb,x_A,x_B) \,\,\, ,
\label{css-W-sol}
\end{equation}
where all large logarithms from $\ln(1/\tb^2)$ to $\ln(M^2)$ have
been completely resummed into the exponential factor
$S(\tb,M)$, and $c$ is a constant of order
unity \cite{CSS-W}. For $\tb>\tb_{max}$,
\begin{eqnarray}
\tilde{W}(\tb,M,x_A,x_B)
=\tilde{W}^{pert}(\tb_{max})
F^{NP}(\tb;\tb_{max})  \,\,\, ,
\label{qz-W-sol-m}
\end{eqnarray}
where the
nonperturbative function $F^{NP}$ is given by
\begin{eqnarray}
F^{NP}
=\exp\bigl\{ -\ln(M^2 \tb_{max}^2/c^2) 
\left[ g_1 \left( (\tb^2)^\alpha - (\tb_{max}^2)^\alpha\right) \right.
 \left.   +g_2 \left(\tb^2 - \tb_{max}^2\right) \right]
-\bar{g}_2 \left(\tb^2 - \tb_{max}^2\right) \bigr\}.
\label{qz-fnp-m}
\end{eqnarray}
Here, $\tb_{max}$ is a parameter to separate 
the perturbatively calculated part from the non-perturbative
input. Unlike in the original CSS formalism,
$\tilde{W}(\tb,M,x_A,x_B)$ is not altered, 
and is independent of 
the nonperturbative parameters  when $\tb < \tb_{max}$. 
In addition, the $\tb$-dependence in 
Eq.~(\ref{qz-fnp-m}) is separated according to different physics
origins. The $(\tb^2)^\alpha$-dependence mimics the
summation of the perturbatively calculable leading power
contributions to the  renormalization group equations
to all orders in the running
coupling constant $\alpha_s(\mu)$. The $\tb^2$-dependence of the
$g_2$ term is a direct consequence of dynamical power corrections to the
renormalization group equations 
and has an explicit dependence on $M$. 
The ${\bar g\ }_{2}$ term represents the effect 
of the non-vanishing intrinsic parton transverse momentum.

A remarkable feature of the $\tb$-space resummation formalism is
that the resummed exponential factor $\exp[S(\tb,M)]$
suppresses the
$\tb$-integral when $\tb$ is larger than $1/M$. 
It can be shown using the saddle point method that, for a large
enough $M$, QCD perturbation theory is valid even at $p\ _T=0$\cite{CSS-W}.
As discussed in Ref.s \cite{qz01,zhang-fai}, the value of the saddle point
strongly depends on the
collision energy \ $\sqrt{s}$, in addition to its well-known $M^2$
dependence. 
The predictive power of the \ $\tb$-space resummation formalism 
should be even better at  the LHC than at the Tevatron.

In $Z^0$ production, since final state interactions are negligible,
power correction can arise only from initial state multiple scattering.
Equations (\ref{qz-W-sol-m}) and (\ref{qz-fnp-m}) represent 
the most general form of $\ \tilde{W}$, and thus (apart from isospin and 
shadowing effects, which will be discussed later), the only way nuclear
modifications associated with scale evolution
enter the $\tilde{W}$ term is through the coefficient 
${g}\ _{2}$. 

The parameters $g_1$ and $\alpha$ of Eq.~(\ref{qz-fnp-m})
are fixed by the requirement of continuity of the function
$\tilde{W}(\tb)$  and its derivative at $\tb=\tb_{max}$.
(The results are insensitive to changes of \ $\tb_{max}$ $\in [0.3$ GeV$^{-1}$,0.7~GeV$^{-1}]$.
We use $\tb_{max}=$ 0.5 GeV$^{-1}$.) 
The value of $g_2$ and ${\bar g}_{2}$ can be obtained by fitting the
low-energy Drell-Yan data.
These data can be fitted with about equal precision if the values 
${\bar g}\ _2=0.25\pm 0.05$ GeV$^2$ and $g_2=0.01\pm 0.005$ GeV$^2$
are taken. As the $\tb$ dependence of the  $g_2$ and ${\bar g}_2$ 
terms in Eq.~(\ref{qz-fnp-m}) is identical, it is convenient to combine
these terms and define
$G_{2}= \ln({M^2 \tb_{max}^2/  {c^2}})g_2 + \bar{g}_2 \,\,\, .$
Using the values of the parameters listed above, we get 
$G_2 = 0.33 \pm 0.07$ GeV$^2$ 
for $Z^0$ production in $pp$ collisions. 
The parameter $G_2$ can be considered 
the only free parameter in the non-perturbative input in Eq. (\ref{qz-fnp-m}),
arising from the power corrections in the renormalization group equations.
An impression about the importance of power corrections can be obtained by
comparing results with the above value of $G\ _2$ 
to those with power corrections turned off ($G_2=0$).
We therefore define the ratio
\begin{equation}
R_{G_2}(p_T) \equiv \left.
\frac{d\sigma^{(G_2)}(p_T)}{dp_T} \right/
\frac{d\sigma(p_T)}{dp_T} \,\,\, .
\label{Sigma-g2}
\end{equation}
The cross sections in the above equation and in the results presented
here have been integrated over rapidity ($-2.4 \leq y \leq 2.4$) 
and invariant mass squared.
For the PDFs, we use the CTEQ5M set\cite{CTEQ5}.

Figure 1 displays the differential cross sections 
and the corresponding $R_{\ G_2}$ ratio 
(with the limiting values of $G_2=0.26$ GeV$^2$
(dashed) and $G_2=0.40$  GeV$^2$ (solid))
for $Z^0$ production as functions of $p\ _T$ at $\sqrt s= 14$ TeV.   
The deviation of $R_{G_2}$ from unity
decreases rapidly as $p\ _T$ increases, and it 
is smaller than one percent for both $\sqrt{s}=5.5$~TeV (not shown)
and $\sqrt{s}=14$ TeV in $pp$ collisions, even when $p\ _T=0$. 
In other words, the effect of power corrections
is very small at the LHC for the whole $p\ _T$  region.

\begin{figure}
\begin{center}
\includegraphics[width=8.0cm]{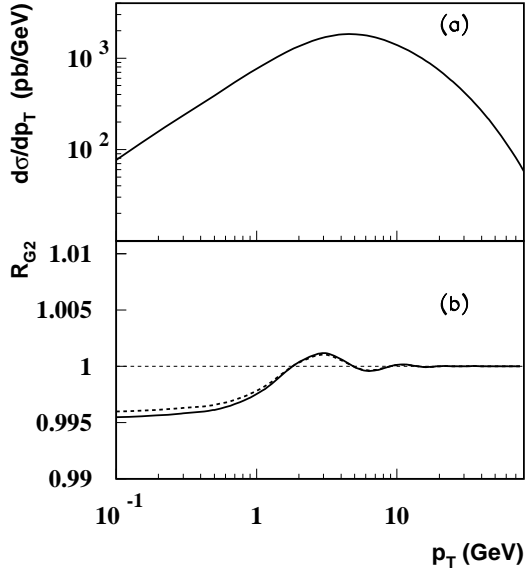} 
\caption{(a) Cross section ${d\sigma / dp_T}$ for $Z^0$ production 
in $pp$ collisions at the LHC with
$\sqrt{s}=14$ TeV; (b) $R_{G_2}$ defined in Eq.~(\protect\ref{Sigma-g2})
with $G_2=$ 0.26 GeV$^2$  (dashed) and 0.40 GeV$^2$  (solid).}
\vspace{-0.3in}
\label{zfig1}
\end{center}
\end{figure}

Without nuclear effects on the hard collision,
the production of heavy vector bosons in nucleus-nucleus ($AB$) collisions
should scale  as the number of hard collisions, $AB$. 
However, there are several additional
nuclear effects on the hard collision in a
heavy-ion reaction. 
First of all, the isospin effect, which come from the difference
between the neutron PDFs and the proton PDFs, is about 2\%  at LHC. 
This is expected, since at the LHC  $x \sim 0.02$,
where the $u-d$ asymmetry is very small. 

The dynamical power corrections entering the parameter $g\ _2$ 
should be enhanced by the nuclear
size, i.e. proportional to $A^{1/3}$. Taking into account the $A$-dependence,
we obtain $G_2 = 1.15 \pm 0.35$ GeV$^2$ for Pb+Pb reactions.
We find that with this larger value of $G_2$, 
the effects of power corrections appear to be enhanced by a factor of  
about three from $pp$ to Pb+Pb collisions at the LHC. 
Thus, even the enhanced power corrections remain
under 1\% when 3 GeV $< p_T < $ 80 GeV. This 
small effect is taken into account in the following nuclear calculations.

Next we turn to the phenomenon of shadowing, 
expected to be a function of $x$,
the scale $\mu$, and of the position in the nucleus. The latter
dependence means that in heavy-ion collisions, shadowing
should be impact parameter ($b$) dependent. 
Here we concentrate on impact-parameter integrated 
results, where the effect of the $b$-dependence of shadowing is relatively 
unimportant\cite{zfpbl02}, and we focus more attention on scale dependence. 
We therefore use EKS98 shadowing\cite{eks} in this work.
We define
\begin{equation}
R_{sh}(p_T) \equiv \left.
\frac{d\sigma^{(sh)}(p_T,Z_A/A,Z_B/B)}{dp_T} \right/
\frac{d\sigma(p_T)}{dp_T} \,\,\, ,
\label{Sigma-sh}
\end{equation}
where $Z_A$ and $Z_B$ are the atomic numbers and $A$ and $B$ are 
the mass numbers of the colliding nuclei, and the cross section
$d\sigma(p_T,Z_A/A,Z_B/B)/dp_T$ has been averaged over $AB$,
while $d\sigma(p_T)/dp_T$ is the $pp$ cross section.
We have seen above that 
shadowing remains  the only significant
effect responsible for nuclear modifications.

\begin{figure}
\begin{center}
\includegraphics[width=8.0cm]{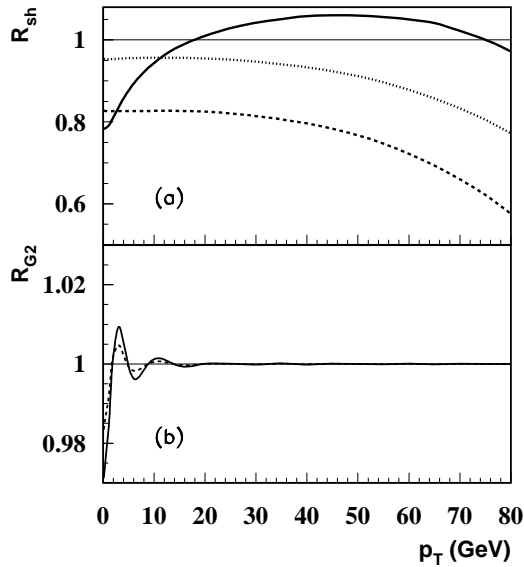} 
\caption{Cross section ratios for $Z^0$ production in Pb+Pb at 
$\sqrt{s}=5.5$ TeV: (a) $R_{sh}$ of Eq.~(\protect\ref{Sigma-sh}) 
(solid line), and $R_{sh}$ with the scale fixed at 
5 GeV (dashed) and 90 GeV (dotted);
(b) $R_{G_2}$ of in Eq.~(\protect\ref{Sigma-g2})
with $G_2=$ 0.8 GeV$^2$ (dashed) and 1.5 GeV$^2$ (solid).}
\label{zfig3}
\vspace{-0.3in}

\end{center}
\end{figure}

In Fig. 2(a),
  $R_{sh}$  (solid line) is surprising, because    
even at $p\ _T=90$ GeV, $x\sim 0.05$, and we are still in the ``strict
shadowing'' region. Therefore, the fact that 
$R_{sh} > 1$  for 20 GeV $< p_T <$ 70 GeV is 
not ``anti-shadowing''.
To better understand the shape of the ratio
as a function of $p\ _T$, 
we also show $R_{sh}$ with the scale fixed at the values 5 GeV (dashed 
line) and 90 GeV (dotted), respectively, in Fig. 2(a). 
The nuclear modification to the PDFs
is only a function of $x$ and flavor in the calculations
represented by the dashed and dotted lines. These two curves
are similar in shape, but rather different 
from the solid line. In $\tb$ space, $\tilde{W}(\tb,M,x_A,x_B)$
is almost equally suppressed
in the whole $\tb$ region if the fixed scale shadowing is used.
However, with scale-dependent shadowing, the 
suppression increases as \ $\tb$ increases, as a
result of the scale $\mu\sim 1/\tb$ in the nPDF. 
We can say that the scale dependence 
re-distributes the shadowing effect. 
In the present case, the re-distribution brings $R_{sh}$ above unity 
for 20 GeV $< p_T <$ 70 GeV. When $p_T$ increases further,
the contribution from the $Y$ term starts to be important, and $R_{sh}$
dips back below one to match the fixed order pQCD result.

We see from Fig. 2 that the
shadowing effects in the $p_T$ distribution of $Z^0$ bosons
at the LHC are intimately related to the scale dependence
of the nPDFs, on which we have
only very limited data\cite{eks}. Theoretical
studies (such as EKS98) are based on the assumption 
that the nPDFs  differ from the
parton distributions in the free proton, but obey the same DGLAP 
evolution\cite{eks}. Therefore, the transverse momentum distribution 
of heavy bosons at the LHC in Pb+Pb collisions can provide a 
further test of our understanding of the nPDFs.

In summary, higher-twist nuclear effects appear to be 
negligible in $Z\ ^{0}$ production at LHC energies. 
We 
have demonstrated that the scale dependence of shadowing 
effects may lead to unexpected phenomenology of shadowing
at these energies. Overall, the $Z\ ^0$ transverse momentum distributions
 can be used as a precision test 
for leading-twist pQCD in the TeV energy region for both,
proton-proton and nuclear collisions. We propose that 
measurements of $Z\ ^{0}$ spectra be of very high priority at the LHC. 

\vspace{0.1in}

\section*{Acknowledgments}


This work was 
supported in part by the U.S. Department of Energy under DE-FG02-86ER-40251.



\end{document}